\documentclass{emulateapj}
\usepackage{natbib, apjfonts}

\shortauthors{KEMPNER, SARAZIN, \& MARKEVITCH}
\shorttitle{{\it CHANDRA}\/ OBSERVATION OF ABELL 2034}

\slugcomment{Accpected for publication in the Astrophysical Journal.}

\begin{document}

\title{Chandra Observation of the Merging Cluster Abell 2034}

\author{Joshua C. Kempner\altaffilmark{1} and Craig L. Sarazin}
\affil{Department of Astronomy, University of Virginia,
P. O. Box 3818, Charlottesville, VA 22903-0818}
\email{jkempner@cfa.harvard.edu,
sarazin@virginia.edu}
\and
\author{Maxim Markevitch}
\affil{Harvard-Smithsonian Center for Astrophysics, 60 Garden St.,
Cambridge, MA 02138}
\email{mmarkevitch@cfa.harvard.edu}
\altaffiltext{1}{Harvard-Smithsonian Center for Astrophysics, 60
Garden St., Cambridge, MA 02138}

\begin{abstract}
We present an analysis of a {\it Chandra}\/ observation of Abell~2034.
The cluster has multiple signatures of an ongoing merger, including a
cold front and probable significant heating of the intracluster medium
above its equilibrium temperature.
We find no evidence for the large cooling rate previously determined for
the cluster, and in fact find it to be roughly isothermal, though with
numerous small-scale inhomogeneities, out to a radius of $\sim$700~kpc.
The cold front appears to be in the process of being disrupted by gas
dynamic instability and perhaps by shock heating.
We find weak evidence for Inverse Compton hard X-ray emission from the
radio relic suspected to be associated with the cold front.
Finally, we study the emission to the south of the cluster, which
was previously thought to be a merging subcluster.
We find no evidence that this ``subcluster'' is interacting with the
main Abell~2034 cluster.
On the other hand, the properties of this region are inconsistent with
an equilibrium cluster at the redshift of Abell~2034 or less.
We explore the possibility that it is a disrupted merging subcluster, but
find its luminosity to be significantly lower than expected for such a
scenario.
We suggest that the emission to the south of the cluster may actually be a
moderate to high redshift ($0.3 \lesssim z \lesssim 1.25$) background cluster
seen in projection against Abell~2034.
\end{abstract}


\section{Introduction}
\label{sec:2034_intro}

Clusters of galaxies are bright X-ray sources, with typical
bolometric luminosities in the range from $10^{43}$ to
$10^{45}$~erg~s$^{-1}$.
In the standard hierarchical model of cluster formation, larger clusters
are formed by hierarchical accretion of smaller clusters.
Occasionally, two clusters of roughly equal mass will merge.
These major mergers are expected to have significant effects on the
intracluster medium (ICM).
Mergers will drive shocks with Mach numbers ${\cal M } \sim 2$,
which can significantly heat the ICM.
While strong, centrally condensed cooling flows are anticorrelated with
major mergers, merger shocks are too weak to completely disrupt cooling
flows at the centers of clusters.
Consequently, cold cores of gas
have been observed to be moving at high velocities through the ICM of
several clusters \citetext{e.g. Abell~2142, Abell~3667;
\citealp{mpn+00,vmm01b}}.
In principle, these thermal effects should all be visible in X-rays: the
shocks as abrupt increases in the entropy of the X-ray emitting gas, and
cold cores as regions of high density, cold gas.

The survival of these cold cores during the merger process has been one
of the most significant discoveries in the area of cluster research made
by {\it Chandra}.
Prior observations did not have the spatial resolution to resolve these
cold cores of gas from the surrounding hotter gas.
The leading edges of these cold cores, dubbed ``cold fronts,'' are
contact discontinuities, where the pressure gradient from the hot gas to
the cold gas is continuous.
The abruptness of the temperature and density change over the small
distance of these fronts means that transport processes (diffusion) across
the fronts must be suppressed \citep{vmm01a,ef00}.

Abell 2034 is a moderate redshift, high X-ray luminosity cluster
at $z = 0.113$ \citep{sr99}.
Previous observations of the cluster with {\it ASCA}\/ and {\it ROSAT}\/
have found, respectively, a cooling flow-corrected temperature of
9.6~keV \citep{whi00} and a bolometric luminosity of $2.2 \times 10^{45}$
erg s$^{-1}$ \citep*{dfj99}.
\citet{whi00} found a best-fit single temperature of 7.6~keV.
Optically, it is quite rich \citep*[ACO richness class 2;][]{aco89}.
It has a cD galaxy which is offset $\sim 1 \arcmin$ from the X-ray
centroid and $2.5 \arcmin$ from the optical centroid of the cluster.
The relative displacements of the cD galaxy, average galaxy position,
and center of the X-ray gas distribution suggest that the cluster is out of
equilibrium.
Indeed, non-equilibrium features were seen in the cluster in a pointed
{\it ROSAT}\/ observation;
the unpublished {\it ROSAT}\/ PSPC image is presented below
(Figure~\ref{fig:2034_rosat}).
It showed an excess of emission to the south of the cluster center which
was suspected to be a merger shock.
Also visible in the {\it ROSAT}\/ image was a surface brightness
discontinuity to the north of the cluster center, opposite the
suspected merger shock.
The detection of a radio relic near the position of the northern
discontinuity \citep{ks01} strengthens the case for the cluster being
out of equilibrium due to a recent or ongoing merger.

\begin{figure*}
\plottwo{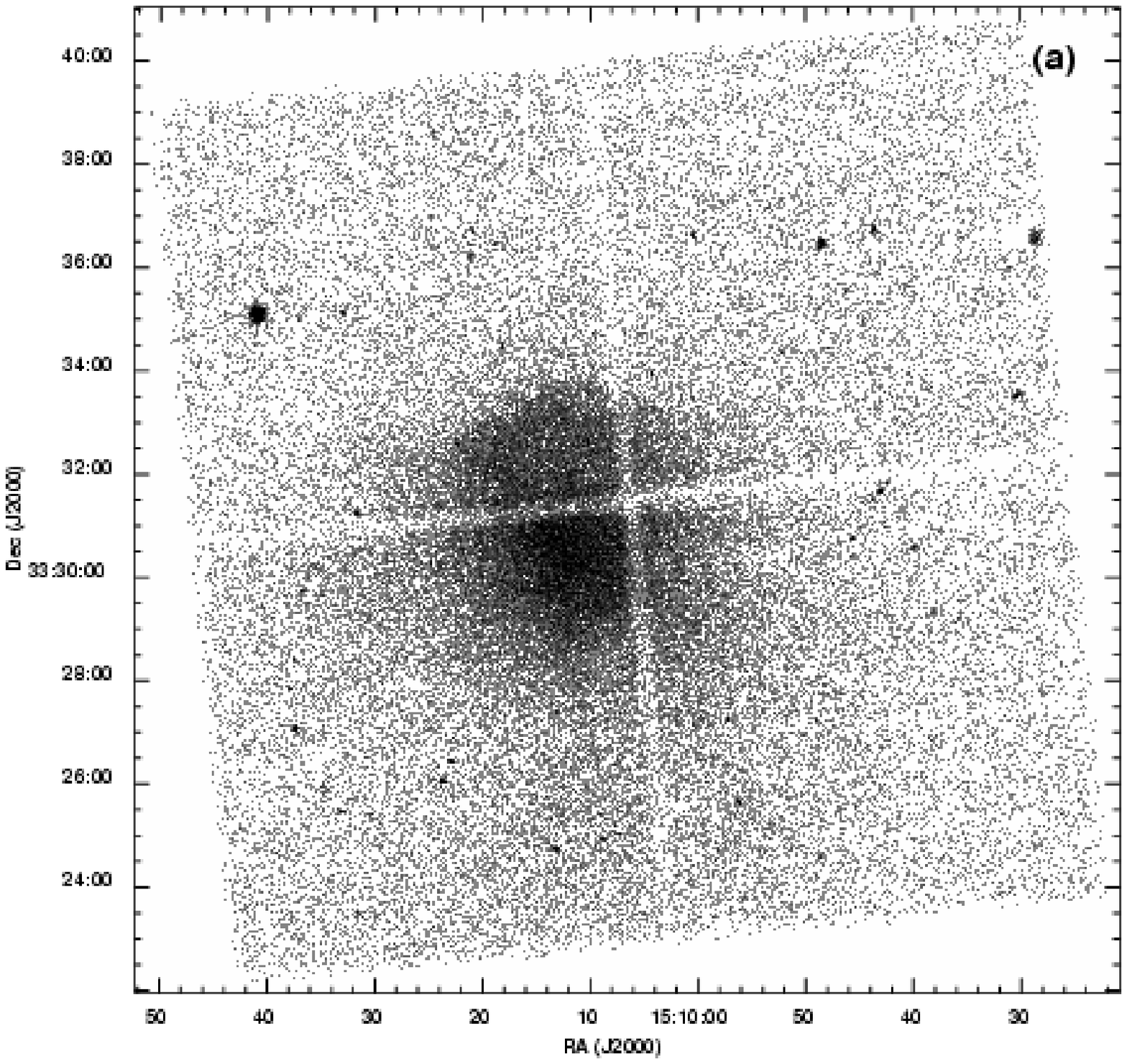}{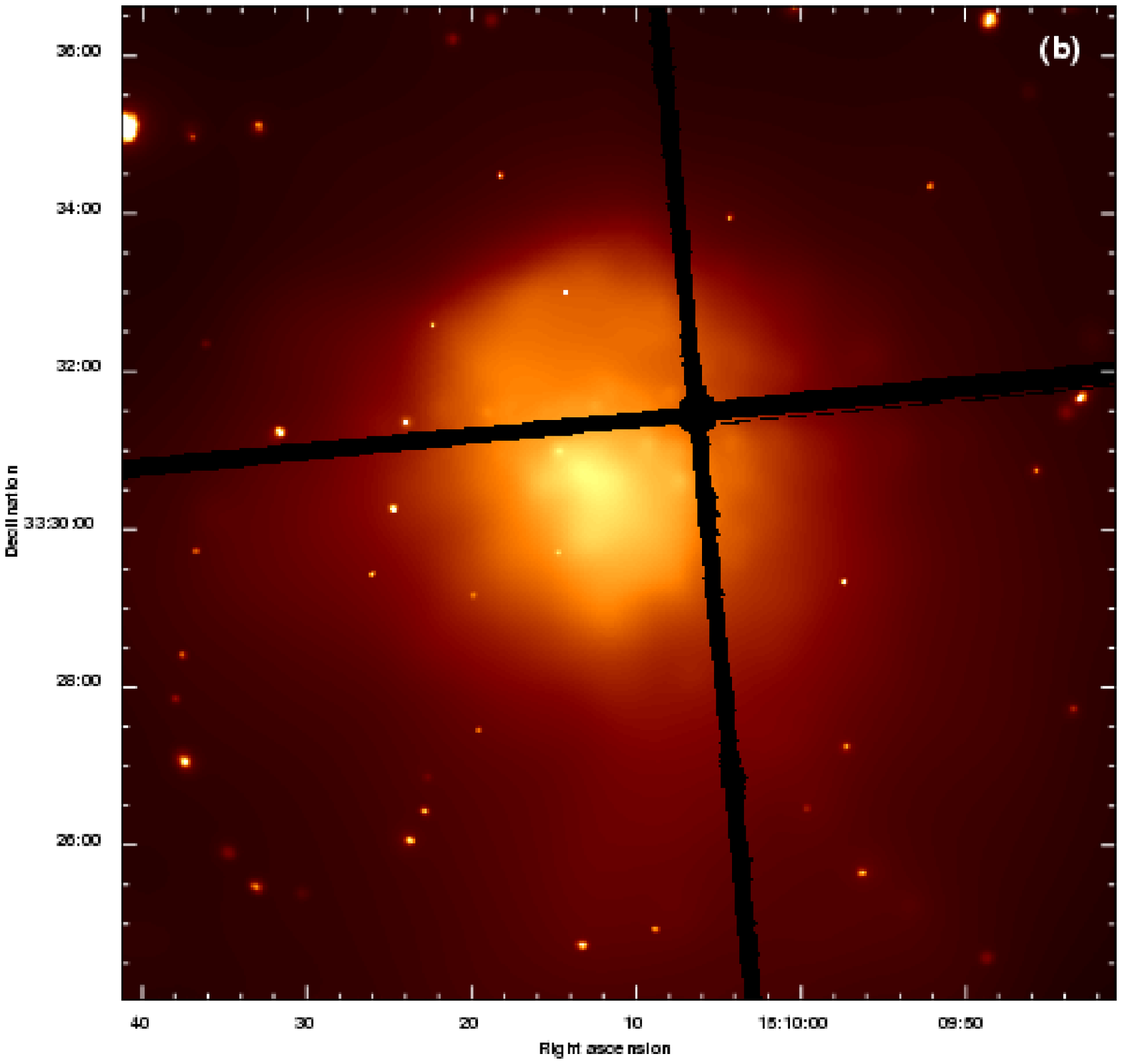}
\figcaption{{\it (a)} Raw X-ray image of Abell~2034 in the 0.3--10 keV
band, uncorrected for background or exposure.
All four ACIS-I chips are shown; the regions of reduced exposure are the
interchip gaps.
This image was binned by a factor of 4, so the pixels in the binned
image are each approximately 2\arcsec$\times$2\arcsec.
{\it (b)} Adaptively smoothed X-ray image of Abell~2034 in the 0.3--10 keV
band, with regions of reduced exposure in the inner chip gaps masked out.
It has been corrected for exposure and for background.
The color is square-root scaled.
The region shown is smaller than that in {\it (a)}, and is centered on the
cluster rather than on the detector.
\label{fig:2034_img}}
\end{figure*}

Despite these pieces of evidence for a merger, Abell~2034 has also been
reported to have a large cooling flow with a cooling rate of
$\sim$90--580~$M_\sun$~yr$^{-1}$
\citep{whi00}.
Usually, large cooling flows are strongly anti-correlated with major
cluster mergers
\citep{bt96}.
Thus, {\it ASCA}\/ and {\it ROSAT}\/ observations did not lead to a
consistent picture of the dynamical state of this cluster.
Higher spatial resolution observations with spatially resolved
spectra and a broad X-ray band (given the high temperature) were
needed.

This makes Abell~2034 an interesting target for a {\it Chandra}\/
observation, particularly since it is located at a sufficiently large
redshift that most of the center of the cluster can be imaged with
the ACIS-I camera.
In this paper, we present the results of such an observation.
All errors are quoted at 90\% confidence unless otherwise stated.
We assume $H_0 = 50$~km~s$^{-1}$~Mpc$^{-1}$ and $q_0 = 0.5$ throughout
this paper.
At the redshift of Abell~2034, $1\arcmin \approx 160$~kpc.

\section{Observation and Data Reduction}
\label{sec:2034_obs}

Abell~2034 was observed on 2001 May 5 in a single 54~ksec exposure.
The data were taken in Very Faint (VF) mode using the four ACIS-I chips
and the S3 chip.
The focal plane temperature was $-$120 C, the frame time was 3.2 s,
and only events with {\it ASCA}\/ grades of 0,2,3,4, and 6 were analyzed.
The S3 chip was used to check the background during the observation;
no background flares were observed.
A binned image, uncorrected for background or exposure, is shown in
Figure~\ref{fig:2034_img}a.

{\it Chandra}\/ data taken in VF mode retain the full 5$\times$5 pixel
event islands, which allow for more thorough rejection of particle
background than do data with only the 3$\times$3 pixel event
islands\footnote{\url{http://cxc.harvard.edu/cal/Links/Acis/acis/Cal_prods/vfbkgrnd/index.html}}.
We applied this additional filtering to our data and to the blank-sky
backgrounds.
We determined the background using the blank sky background files
included in the {\it Chandra}\/ Calibration
Database\footnote{\url{http://asc.harvard.edu/caldb/}}.
We used the period D background files, which include data taken in VF mode.

The ancillary response files (arfs) used in the spectral analysis were
corrected for the reduction in transmission of the CCD window at low
energies using the ``apply\_acisabs'' tool in CIAO.  At the time of the
observation, the reduction in sensitivity due to this effect was about 20\%
at 0.8 keV.  Due to additional uncertainty in the calibration at low
energies, we restricted our analysis to the range 0.8--9.0 keV.

\section{X-ray Image}
\label{sec:2034_image}

The raw {\it Chandra}\/ image, binned by a factor of 4 and uncorrected for
background or exposure, is shown in Figure~\ref{fig:2034_img}a.
At its moderate redshift, Abell~2034 fills much of the ACIS-I
detector on {\it Chandra}.
The cluster was positioned on the detector so that the aimpoint would be
about halfway between the cD galaxy and the centroid of the X-ray
emission as determined from the {\it ROSAT}\/ image.
The image shows several interesting features in the cluster itself, and
a number of point sources.

A smoothed image of the diffuse emission is shown in
Figure~\ref{fig:2034_img}b.
The image was adaptively smoothed using the {\it csmooth}\/
tool from the CIAO software
package\footnote{\url{http://asc.harvard.edu/ciao/}}.
Blank sky background and exposure images were smoothed using the same
kernel, and were then used to correct the smoothed raw image.

\begin{figure}
\plotone{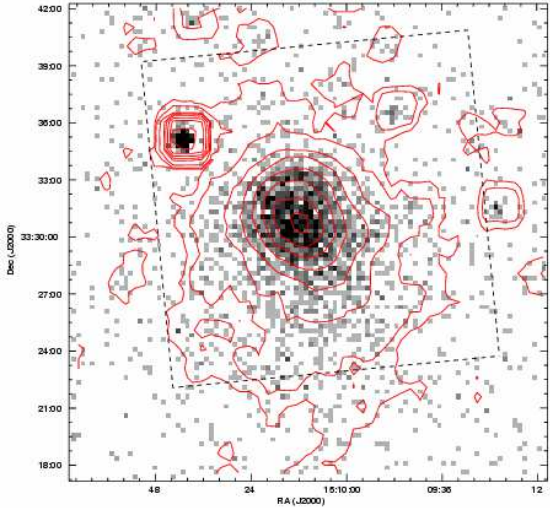}
\figcaption{{\it ROSAT}\/ PSPC image of Abell~2034 in the 0.5--2.4 keV band,
corrected for background and exposure.
The contours show levels of constant surface brightness, with
square-root spacing.
The dashed line shows the approximate field of view of the ACIS-I.
\label{fig:2034_rosat}}
\end{figure}

As can be seen in Figures~\ref{fig:2034_img}a and \ref{fig:2034_img}b,
the cluster has significant
emission out to a distance of $5\arcmin$--$6\arcmin$ in all directions.
There is a sharp discontinuity in the surface brightness
$\sim$$3\arcmin$ to the north of the cluster center.
We will refer to this feature as the ``northern cold front.''
It is discussed in detail below (\S\ref{sec:2034_cold_front}).
This feature was seen in the earlier {\it ROSAT}\/ image
(Figure~\ref{fig:2034_rosat}; the image was corrected for background and
exposure using the {\it ROSAT}\/ extended object analysis tools by S.
Snowden\footnote{\url{ftp://legacy.gsfc.nasa.gov/rosat/software/fortran/sxrb/}}).
This feature was one of the motivations for observing this cluster
with {\it Chandra}.
Based on the {\it ROSAT}\/ observations, which provided no useful
information on the X-ray spectra near this feature, we had interpreted
this feature as a possible merger shock.

Also visible in the raw and smoothed X-ray images is an excess of
emission to the south of the cluster center.
We will refer to this region as the ``south excess;''
it is discussed in more detail below (\S~\ref{sec:2034_south}).
This feature was also seen in the earlier {\it ROSAT}\/ image
(Figure~\ref{fig:2034_rosat}), where it appeared as a rather distinctive
arc of emission running from the southeast to the northwest.
Initially, we suspected that this feature might be a merger shock
cutting across the cluster.
The extensions of this arc in the {\it ROSAT}\/ image are due, at least
partially, to unresolved point sources seen in the {\it Chandra}\/ image.
Because {\it Chandra}\/ is able to better resolve these distinct regions
of emission and separate out point sources, the south excess appears as
a fairly regular region of diffuse emission rather than as the arc which
we believed to see in the {\it ROSAT}\/ image
(Figure~\ref{fig:2034_rosat}).

\begin{figure}
\plotone{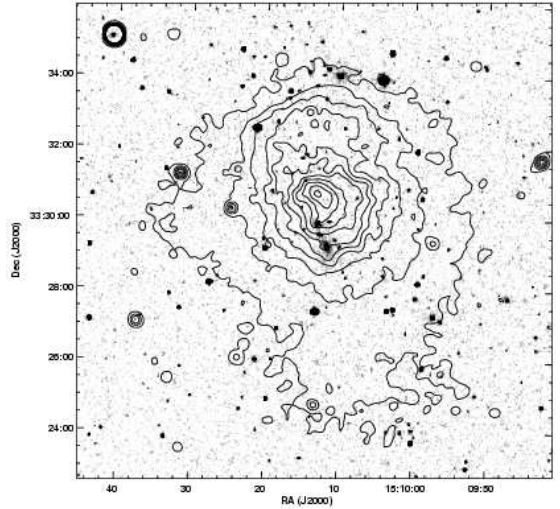}
\figcaption{Digital Sky Survey II (DSS2) R-band optical image of the
same region as shown in Figures~\protect\ref{fig:2034_img}a and
\protect\ref{fig:2034_img}b.
Contours of the {\it Chandra}\/ X-ray image are superposed.
The surface brightness contours are linearly spaced, ranging from
$1.6\times10^{-5}$ to $1.8\times10^{-4}$~counts s$^{-1}$ pixel$^{-1}$,
and are smoothed on a scale of 16 pixels.
Some distortion of the contours is visible along the ACIS-I chip gaps.
\label{fig:2034_optical}}
\end{figure}

The cluster does not have a particularly sharply peaked core, suggesting
that it does not contain a strong cooling flow.
The core of the cluster has a somewhat irregular structure.
The central arcminute or so is shaped like a curved teardrop, with the
brightness peak centered on the broad end of the teardrop, to the north
of center.
Outside of 1\arcmin, the cluster becomes more regular and circular.
This structure is clearly visible in the contours in
Figure~\ref{fig:2034_optical}.

An optical image of the Digital Sky Survey (DSS) is shown in
Figure~\ref{fig:2034_optical}.
Note that the cD galaxy which is the brightest cluster member is not
located at the peak in the X-ray surface brightness, but is
$\sim$77\arcsec\ south of the peak.
There is a smaller peak in the X-ray surface brightness which does
coincide with the central cD galaxy.

A very bright cluster galaxy, also classified as a cD, is located
NNW of the X-ray peak near the location of the northern cold front.
There are several other bright elliptical galaxies in this
neighborhood, and numerous fainter galaxies.
In fact, the number of moderately bright to faint galaxies in this
region significantly outnumbers that in the region of the more central
cD galaxy.
The majority of these galaxies are north, or ahead, of the cold front,
perhaps suggesting that the collisionless galaxies have moved ahead of
the collisional gas, which has been slowed during the merger.
Just inside the cold front, there is a significant excess of X-ray emission
relative to that expected from a smooth $\beta$-model, as we discuss later
in \S\ref{ssec:2034_global_surf}.

There are no bright galaxies and no obvious excess in the galaxy
density evident in the DSS image of the region covered by the
south excess.
Two faint galaxies ($m_{\rm R} \sim 17$) are visible on the western edge of
the south excess, and are associated with the wide angle tail
(WAT) radio galaxy FIRST~J150957.2+332716 and the narrow angle
tail (NAT) radio galaxy FIRST~J150959.5+332746.
The latter is also visible as an X-ray source in the {\it Chandra}\/ image.

\section{Global Cluster Properties}
\label{sec:2034_global}

\subsection{Radial Gas Distribution}
\label{ssec:2034_global_surf}

The X-ray image of Abell~2034 shows a number of strong features
which suggest that the cluster may not be relaxed.
These include the northern cold front, the south excess,
the lumpy structure near the center, and the displacement of the
central cD galaxy from the X-ray peak.
Despite these local irregularities, the radial surface brightness
profile of the cluster is fairly consistent over most azimuthal angles.

\begin{figure*}
\plottwo{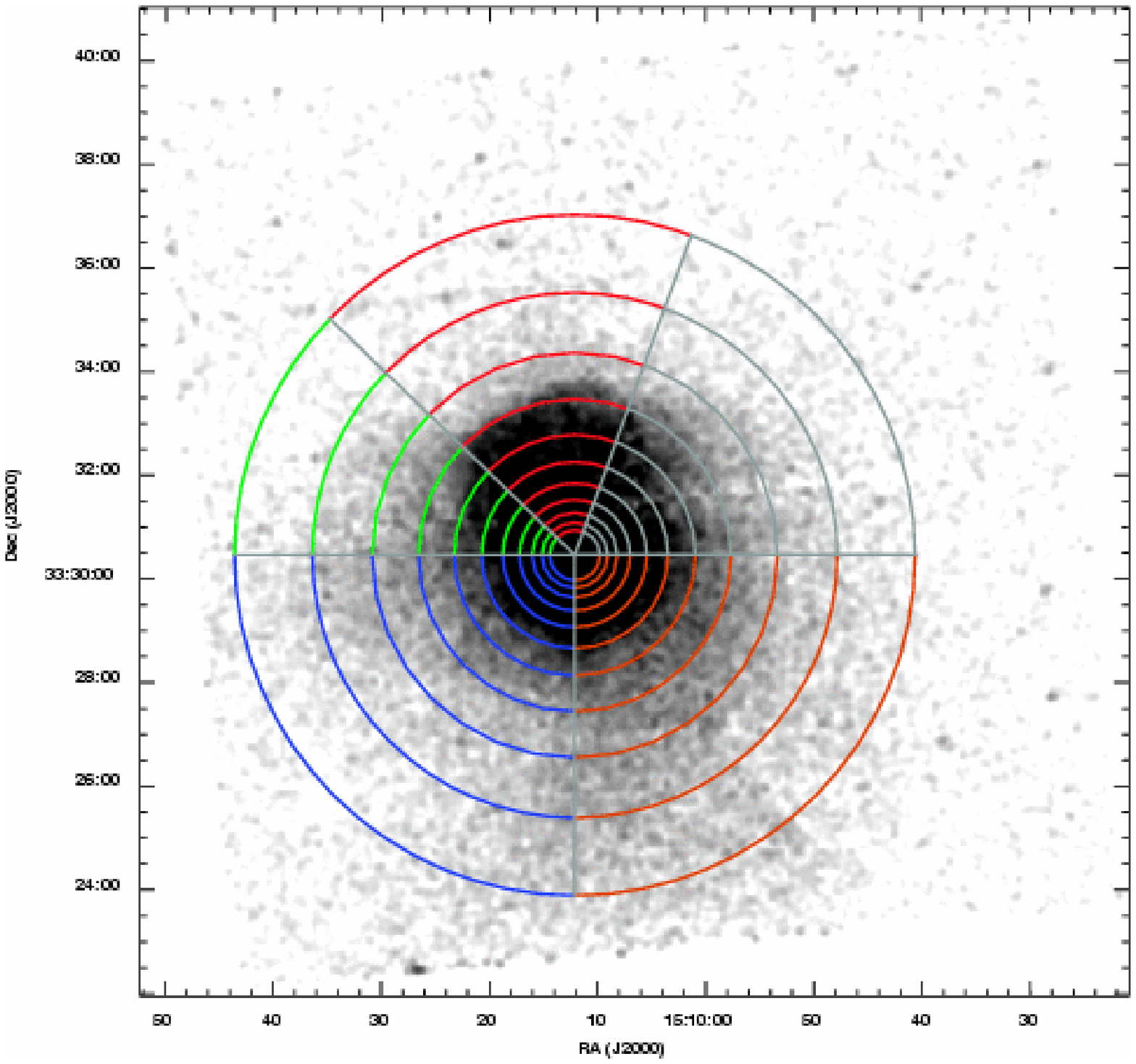}{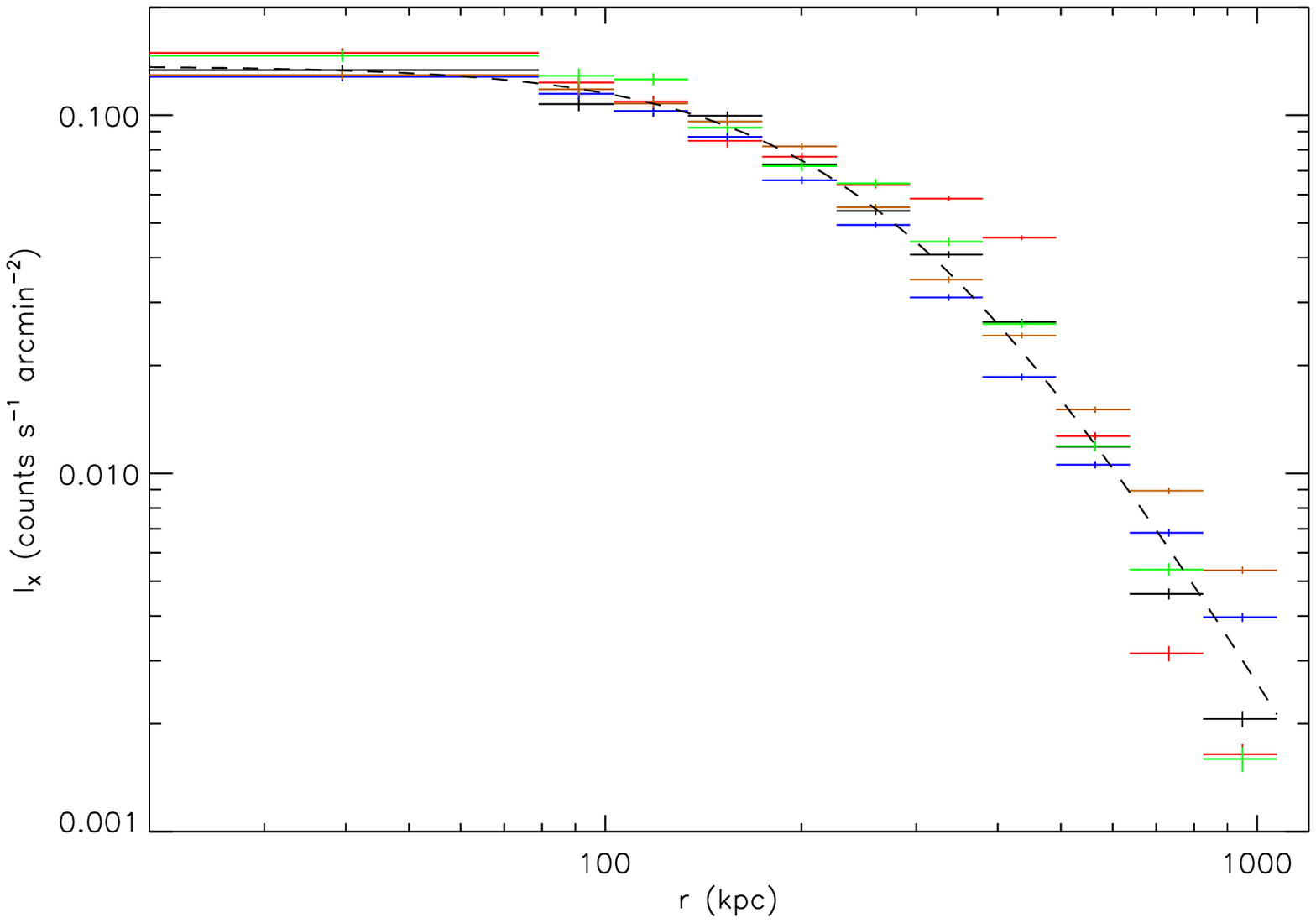}
\figcaption{{\it Left panel:} Gaussian smoothed X-ray image of Abell~2034
overlaid with the sectors used for fitting radial surface brightness and
temperature profiles.
All point sources have been removed.
{\it Right panel:} Surface brightness profiles in the five sectors.
The dashed line is the best fit $\beta$-model from a simultaneous fit to
all five profiles.
The data points correspond to the sectors in the left panel with matching
colors, with the exception of the black points, which are displayed in grey
in the left panel for clarity.
The $y$-axis error bars have been shifted slightly from the bin centers
so as not to obscure each other; the black points are unshifted.
\label{fig:2034_global_prof}}
\end{figure*}

We determined the radial surface brightness profile in 5 sectors
where the cluster is approximately circularly symmetric within each sector.
The sectors and annuli used to extract the surface brightness profile
are shown in Figure~\ref{fig:2034_global_prof}a.
The annuli were selected to have between 1000 and 2000 source counts per
annulus in order to produce reasonable spectral fits.
This meant that for the purposes of the surface brightness analysis
presented here they had a much higher flux than was necessary.
In Figure~\ref{fig:2034_global_prof}b, we show the radial surface
brightness profiles in these five sectors.
The sectors' common origin is the approximate centroid of the cluster
emission.
A $\beta$-model fit to the sum of these profiles yields the best-fit
core radius of $r_c = 290\pm10$~kpc and slope of $\beta = 0.69\pm0.02$
(1-$\sigma$ errors).
We could not find any previously published values for these parameters for
comparison.
This model provides a reasonably good fit to all of the inner regions of
the sectors except
for the one encompassing the northern edge, plotted in red in
Figure~\ref{fig:2034_global_prof}b.
This particular sector shows an excess of emission immediately inside
the edge and a smaller deficit outside, but at smaller and larger radii
is reasonably consistent with the profiles in the other sectors.

\begin{figure}
\plotone{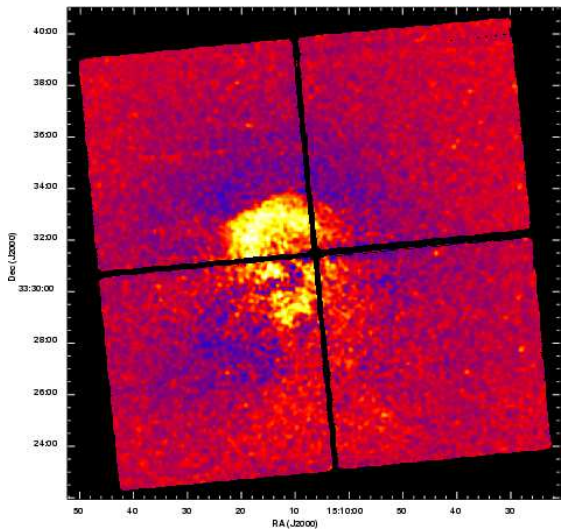}
\figcaption{Residuals after subtraction of the $\beta$-model which provides
the best fit to the blue sector only.
The $\beta$-model used has $r_c = 190$ kpc and $\beta = 0.54$.
The image has been smoothed by a 2\arcsec\ Gaussian.
\label{fig:2034_beta_resid}}
\end{figure}

The southwest sector has a noticeably flatter profile with brighter
emission at large radii due to the south excess region.
The south excess may in fact be a separate, possibly more distant
cluster seen in projection against Abell~2034, as we discuss in detail
in \S\ref{sec:2034_south}.
In Figure~\ref{fig:2034_beta_resid}, we show the residuals after
subtracting a $\beta$-model from the Gaussian smoothed image shown in
Figure~\ref{fig:2034_global_prof}a.
Since the southeast sector is largely free of emission from the south
excess and from other substructure within the cluster, we fit a
$\beta$-model separately to this sector for use in creating
Figure~\ref{fig:2034_beta_resid}.
The resultant $\beta$-model fit has $r_c = 190$ kpc and $\beta = 0.54$,
which is a significantly smaller core radius and flatter slope than the
best fit to all the sectors including the south excess and internal
substructure.
As is evident in Figure~\ref{fig:2034_beta_resid}, however, this still
oversubtracts slightly the large-scale diffuse emission in the cluster,
particularly at radii less than $\sim$800 kpc.
This oversubtraction is probably due to the $\beta$-model still being
affected by a small contribution from the south excess and from the excess
at the very center of the cluster.

\subsection{Spectral Properties}
\label{ssec:2034_global_spec}

To test the previous report using data from {\it ASCA}\/ of a large cooling
flow in Abell~2034 \citep{whi00}, we extracted a spectrum from a circular
region with radius equal to the core radius determined in
\S\ref{ssec:2034_global_surf}.
We fit it with a single temperature MEKAL model \citep*{kaa92,log95}, and
then with a single temperature plus a cooling flow.
In both cases we fixed the absorption to the Galactic value of $1.58
\times 10^{20}$~cm$^{-2}$ \citep{dl90}.
The single temperature fit determined a temperature of $7.9 \pm 0.4$~keV,
and an abundance of $0.29 \pm 0.07$ times solar, with $\chi^2 / d.o.f. =
1.4$.
This abundance is consistent to within the errors with that determined by
\citet{whi00} using data from {\it ASCA}.
The addition of a cooling flow component, with the minimum temperature set
to essentially zero and the maximum temperature and abundance tied to the
values of the single-temperature component, did not improve the fit.
The resultant cooling rate to low temperature was determined to be
$23^{+21}_{-20}~M_\sun$~yr$^{-1}$.
With the low temperature as a free parameter, the fit was too poorly contrained
to provide any useful limits.
In short, we find evidence for only a small amount of cooling in the center
of the cluster.
It is likely that the large cooling rate measured by \citet{whi00} was
due to the presence of cooler gas in the cold front
(see \S\ref{sec:2034_cold_front}) and in the south excess (see
\S\ref{sec:2034_south}) rather than in a classical cooling flow in the
cluster core.

\begin{figure}
\plotone{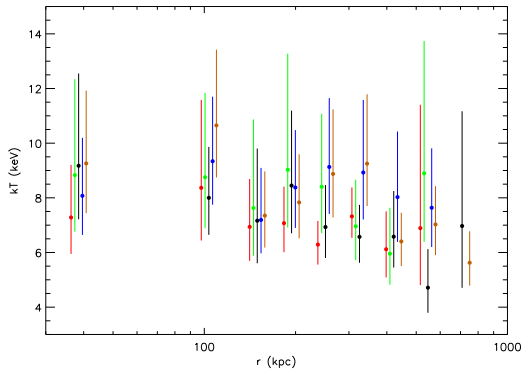}
\figcaption{Temperatures from single temperature fits to concentric annular
wedges with abundance fixed at $0.25\, Z_\sun$.
The colors of the data points correspond to the same sectors as in
Figure~\ref{fig:2034_global_prof}.
There were not enough source counts to do the spectral analysis in the
outer most annulus in 3 of the five sectors.
The points have been shifted slightly from the bin centers so as not to
obscure each other; the black points are unshifted.
\label{fig:2034_global_Tprof}}
\end{figure}

We extracted spectra in the same sectors as were used in
\S\ref{ssec:2034_global_surf}, and fit them with a single temperature
MEKAL model.
For the initial spectral fits, we fixed the absorption at the Galactic
value.
We first allowed only the temperature and normalization to vary, with the
abundance fixed at the mean value of 0.29 times solar determined above.
We then allowed the abundance to vary and re-fit the spectra.
Allowing the abundances to vary did not change the temperatures we measured
by a significant amount, and the abundances we determined were not well
constrained---in most cases they provided only upper limits.
There is no evidence from these spectra, however, for any large abundance
gradients across the cluster, and all the abundances we measured were
consistent with the value of 0.29 times solar determined above.
We therefore use the fits with fixed abundance in our subsequent analysis.
The results of these fits are shown in Figure~\ref{fig:2034_global_Tprof}.

With the exception of the southwest quadrant, where contamination from
the south excess is significant at large radii, we find no evidence
for a large scale temperature gradient like that seen in some clusters
\citep{mfs+98}.
The temperature gradients reported by \citet{mfs+98}, however, were
generally at radii $\sim$1~Mpc, whereas our temperature profiles are
only significant out to $\sim$700~kpc.

\subsection{Temperature Structure}
\label{sec:2034_tmap}

\begin{figure}
\plotone{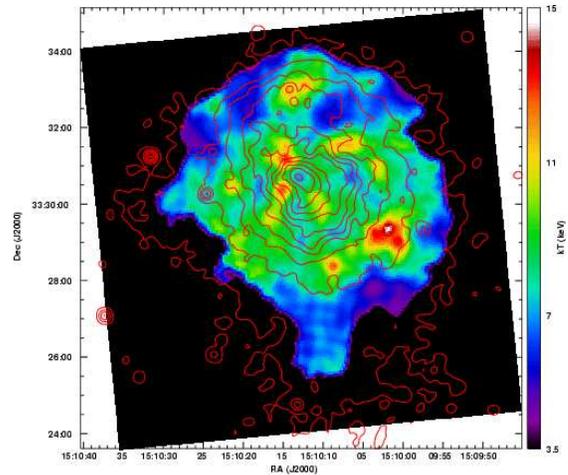}
\figcaption{Adaptively binned temperature map overlaid with surface
brightness contours.
The image has been smoothed with a $\sigma = 1.5$ pixel Gaussian to
eliminate statistically insignificant pixel-to-pixel variations.
The contours are the same as in Figure~\ref{fig:2034_optical}.
\label{fig:2034_tmap}}
\end{figure}

In Figure~\ref{fig:2034_tmap}, we present an adaptively binned temperature
map of the cluster.
The temperature map was created using the adaptive binning algorithm
described in \citet{hd00} and \citet*{hwd02}.
The algorithm extracts a spectrum at each pixel in the output map from a
region which is as small as possible while containing at least 800 counts.
The maximum extraction region was 63\arcsec$\times$63\arcsec.
The smallest extraction region used was 26\farcs6$\times$26\farcs6.
For pixels around which it was not possible to extract 800 counts from a
region of this maximum size, no spectrum was extracted and no temperature
was calculated.
Each spectrum was corrected for background using a spectrum from a matching
region in the blank sky background file discussed in \S\ref{sec:2034_obs}.
Regions containing point sources were eliminated from the event lists used
for the source and background spectra.
The spectra were grouped to have a minimum of 20 photons per bin, then fit
by a single MEKAL model with absorption fixed at the Galactic value, and
abundance fixed at $0.29\, Z_\sun$.
The negative errors in the temperature map are less than 15\% everywhere.
The positive errors are larger since the average temperature is close to or
above the upper limit of the {\it Chandra}\/ spectral band, i.e. $\gtrsim 9$
keV.
Where the temperature is less than $\sim$8 keV, the positive errors are
between 15\% and 30\%.

The temperature map shows inhomogeneities on scales roughly the size of the
binning scale and larger.
Since the binning washes out any inhomogeneities on smaller scales, and since
the inhomogeneities span a range of scales up to several times the binning
scale, it is likely that better photon statistics would reveal variations on
even smaller scales.
The temperature variations show no correlation with surface brightness.
Because of the overall high temperature of the gas relative to the useful
upper limit of the Chandra energy range, the excursions in the map to lower
temperatures from the ambient temperature of $\sim$9 keV are statistically
significant, while the excursions to higher temperature are not.
Thus, the hot spot to the southwest is not significantly hotter than the
surrounding gas at 90\% confidence.
In contrast, the cold spot corresponding to the central brightness peak is
cooler than the surrounding gas at 90\% confidence.
On the other hand, excursions to higher temperature in regions of lower
ambient temperature, such as the extended hot region just inside the
northern cold front, are significant.
This hot region is coincident with the middle part of the cold front, which
has a less well defined edge than the rest of the front.
As we will discuss in \S\ref{sec:2034_cold_front}, this feature may be
related to the development of a hydrodynamic instability in the front.
A detailed spectral analysis of this hot region shows that it is
significantly hotter than the gas to either side of it:
$9.9^{+3.4}_{-1.9}$~keV compared to $6.1^{+1.2}_{-0.9}$~keV elsewhere just
inside the cold front for a single-temperature MEKAL model with the
absorption and abundance fixed as above.
We discuss this hot spot in greater detail in \S\ref{sec:2034_cold_front}

\subsection{Mass Distribution}
\label{ssec:2034_mass}

\begin{figure}
\plotone{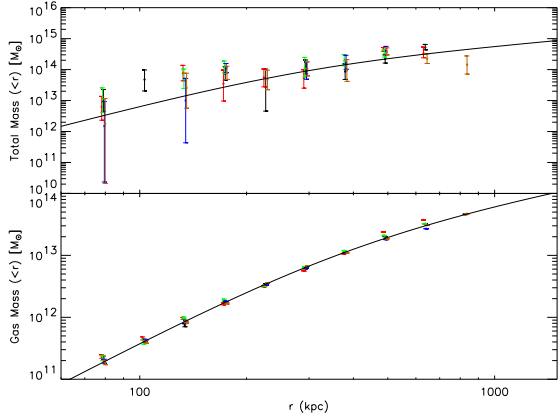}
\figcaption{Gravitational mass (top panel) and gas mass (bottom panel)
interior to the radius at which the mass is plotted.
The colors correspond to the same sectors as in the previous figures.
Where the pressure in a given annulus was found to be greater than the
pressure in the annulus immediately interior to it, no data are plotted.
The points have been shifted slightly from the bin centers so as not to
obscure each other; the black points are unshifted.
At radii where the pressure difference would imply a negative mass, no
points are plotted.
The mass profiles derived from the $\beta$-model are overplotted as solid
curves.
\label{fig:2034_mass}}
\end{figure}

Combining the surface brightness and temperature profiles from above, we
produce two mass profiles of the cluster in each of the five sectors:
a profile of the gas mass, and a profile of the total gravitational
mass.
The gas mass is determined primarily by the surface brightness profile,
which is
deprojected under the assumption that the cluster is spherically
symmetric.
The temperatures from the variable-abundance fits were used to derive
the emissivity.
The gravitational mass is calculated by deriving the pressure in each
concentric annulus and applying the condition of hydrostatic equilibrium
to derive the necessary mass interior to the given shell.
The gravitational and gas masses are plotted in
Figure~\ref{fig:2034_mass}.

As can be seen from the figure, the gas masses determined from the different
sectors are quite consistent.
This should not be surprising given the similarities between the surface
brightness profiles.
The gravitational mass profiles are not as well determined as those of
the gas mass, for several reasons.
The first is the scatter in the temperature measurements, which
contribute directly to the pressure calculations.
The second is that the calculation of the gravitational mass assumes that
the cluster is in hydrostatic equilibrium.
Since it is undergoing a merger and shows evidence of relative motion
between the gas in different parts of the cluster, such as the cold
front in sector 2 (the red points in Figure~\ref{fig:2034_mass}), it is
almost certainly {\em not} in hydrostatic equilibrium.
Not only is the gas unlikely to be at rest relative to the gravitational
potential, it has also probably been heated by the merger beyond the
temperature one would derive simply by applying the hydrostatic
condition to a cluster with the combined mass of the original
subclusters \citep*{rsr02}.

Also shown in Figure~\ref{fig:2034_mass} is a mass profile derived from
the $\beta$-model fit produced in \S\ref{ssec:2034_global_surf}, under
the assumption that the cluster is isothermal.
Here we used the mean temperature from the spectral models in
\S\ref{ssec:2034_global_spec} with fixed abundance.
This model fits the data remarkably well, both for the gas mass and
total mass profiles, particularly at large radii.
At radii less than the core radius of the $\beta$-model ($r<290$ kpc),
the fit to the gravitational mass is worse.
In this region, however, the pressure gradient should be small, and
since our method for determining the gravitational mass from the data
involves computing pressure differences between points, the accuracy of
mass at these radii is suspect.

Using the outermost bin for which we have estimates of both the gas mass
and the gravitational mass, the gas mass fraction is $\sim$13\%.  The
$\beta$-model mass profile gives a similar fraction.  This fraction is
consistent with the typical value of 10--20\% \citep[][all
for $H_0 = 50$~km~s$^{-1}$~Mpc$^{-1}$]{djf95, asf02}.

\section{South Excess}
\label{sec:2034_south}

As mentioned above, there is a faint, large scale excess of emission to
the south and southwest on the cluster periphery.
In the {\it ROSAT}\/ image of the cluster (Figure~\ref{fig:2034_rosat},
this excess appeared to be linear, leading us to believe that it was
likely to be a merger shock.
As mentioned in \S\ref{sec:2034_image}, the {\it Chandra}\/ image of the
cluster shows this region to be much more diffuse, and not a sharp
feature at all.
It appears much more round, and probably extends off the south edge of the
detector.
The {\it ROSAT}\/ image, with its large field-of-view, shows that the south
excess does indeed extend about 4\arcmin\ past the edge of the ACIS-I
detector.
The excess is slightly more than 8\arcmin\ across, so approximately half of
the south excess is outside the field-of-view of the ACIS-I detector.

We fit the south excess with a two-temperature MEKAL model, with one
component's parameters fixed at the values determined at the same distance
from the cluster center and to the northeast of the excess, and the other
component's temperature, abundance and normalization allowed to vary.
This second component was found to have a temperature of
$4.9^{+1.2}_{-0.8}$~keV and an abundance of $< 0.73 \, Z_\sun$.
Thus, the south excess is cooler than the gas in the main body
of the Abell~2034 cluster.
The excess in this region therefore cannot be due to shock compression
of this hotter gas.

One possible interpretation is that the south excess is a smaller cluster
which is merging with Abell~2034, but which has not yet been shock heated
by the interaction.
This would explain the low temperature of the south excess.
However, the flux from the south excess is much too low for a
$\sim$4.9~keV cluster at $z=0.113$, if it follows the standard
$L_X$--$T$ relation for clusters \citep[e.g.][]{mar98}.

Another possible explanation of the south excess is that it was originally
a relatively small, cool, and X-ray faint subcluster which has been
shock heated to $kT = 4.9$ keV as a result of a merger with Abell~2034.
There are two possible concerns with this explanation.
First, there is no evidence for a strong interaction between the
south excess and the main cluster in Abell~2034.
Second, the observed temperature of the south excess is lower than
might be expected for gas shocked while merging with the main
cluster, given the very high temperature of Abell~2034.
This is true even for the measured upper limit on the temperature of the
south excess.
However, it is possible that the merger shock was a weak oblique shock
resulting from an offset merger (a merger with a large impact
parameter), or that the south excess region originated as substructure
within Abell~2034 and did not fall through the full potential well of
the cluster.
It is also possible that while the subcluster was initially shock heated,
it has since had time to adiabatically expand as the center of its dark
matter potential has moved ahead of the gas.
This would require the subcluster either to have a very small mass or to be
a quite advanced merger, probably having already fallen through the
cluster.
Otherwise, the cumulative effect of ram pressure over the lifetime of the
merger would be insufficient to decouple the gas from the potential.

\begin{figure*}[ht]
\plottwo{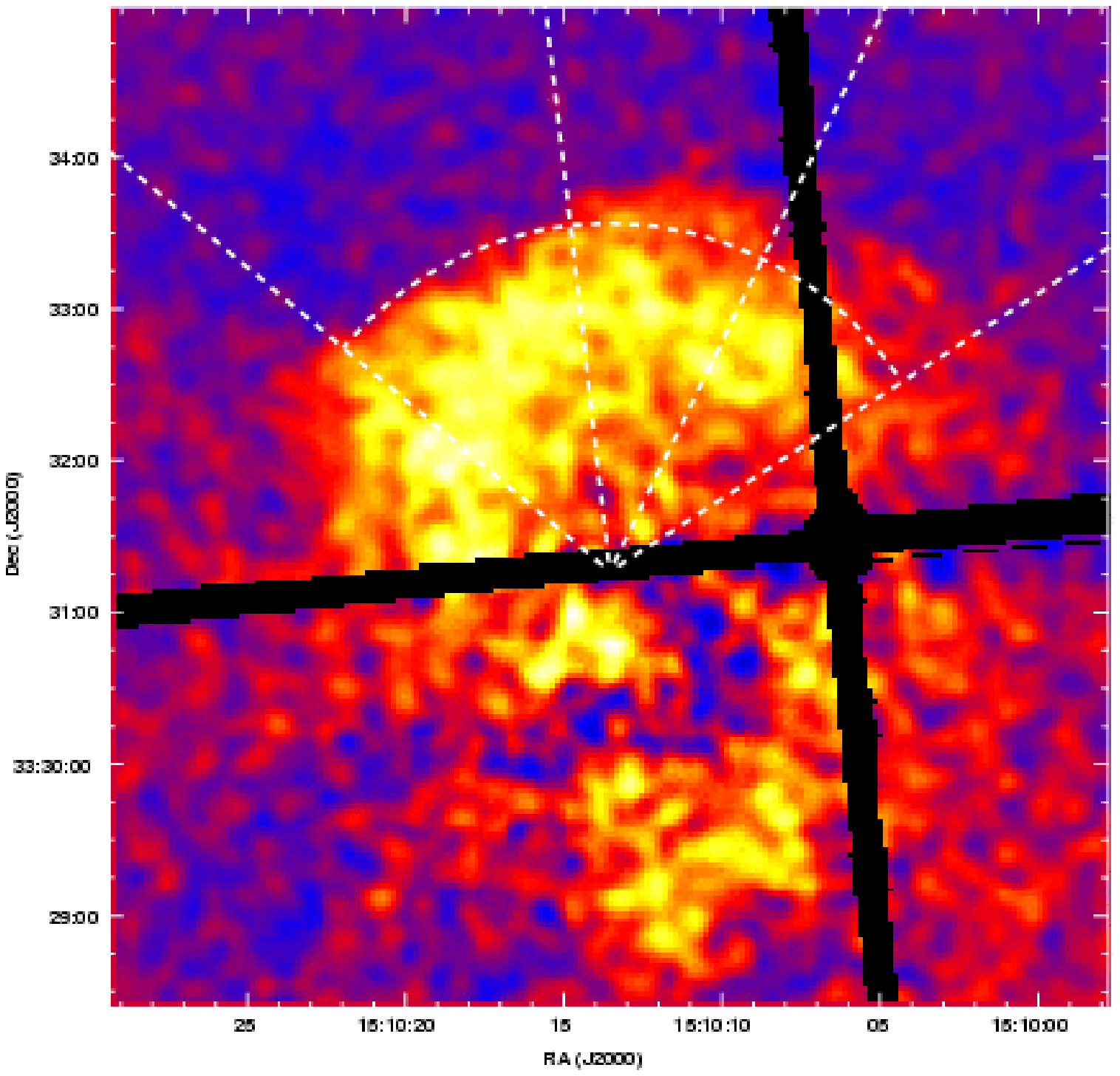}{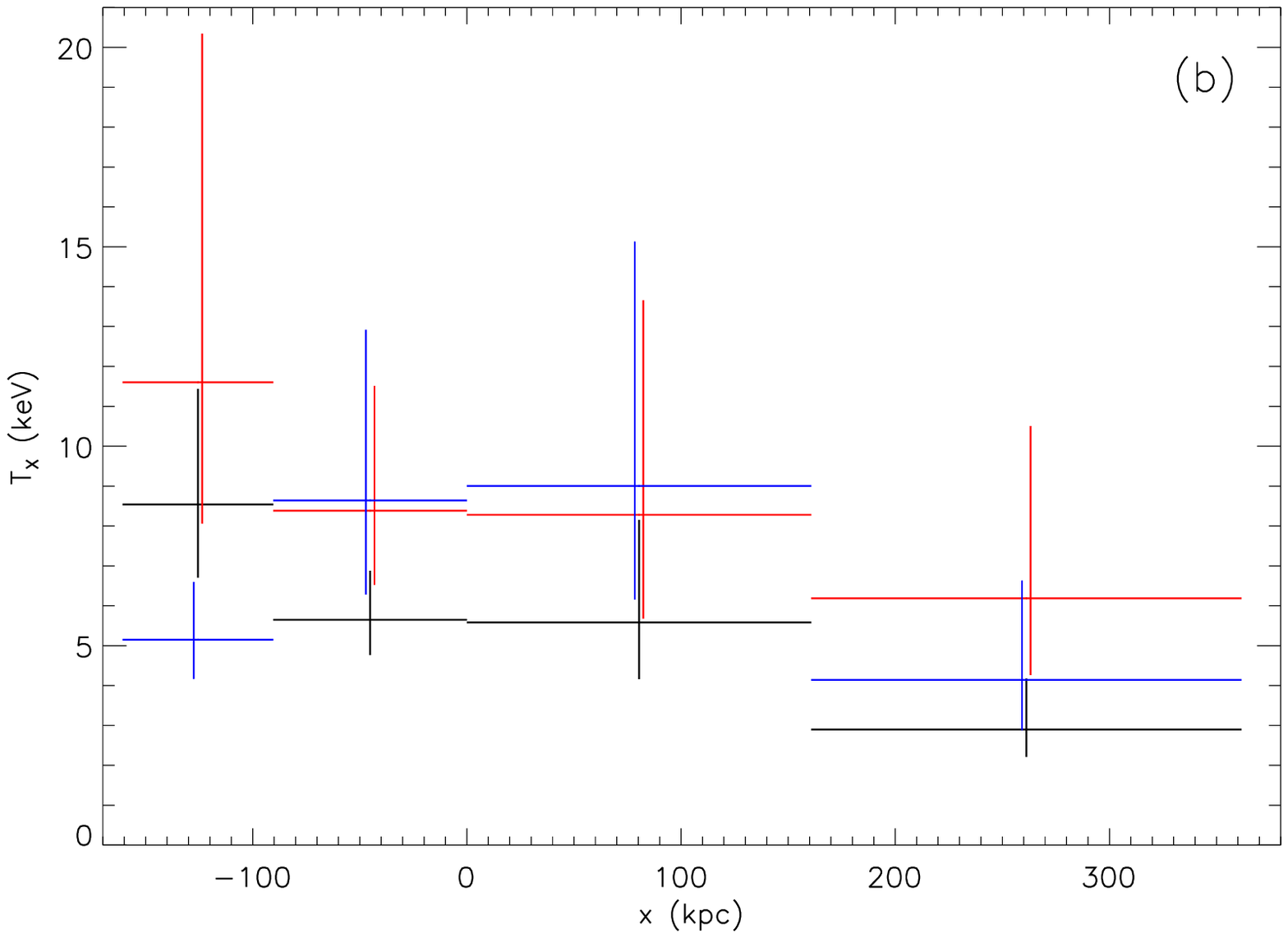}
\plottwo{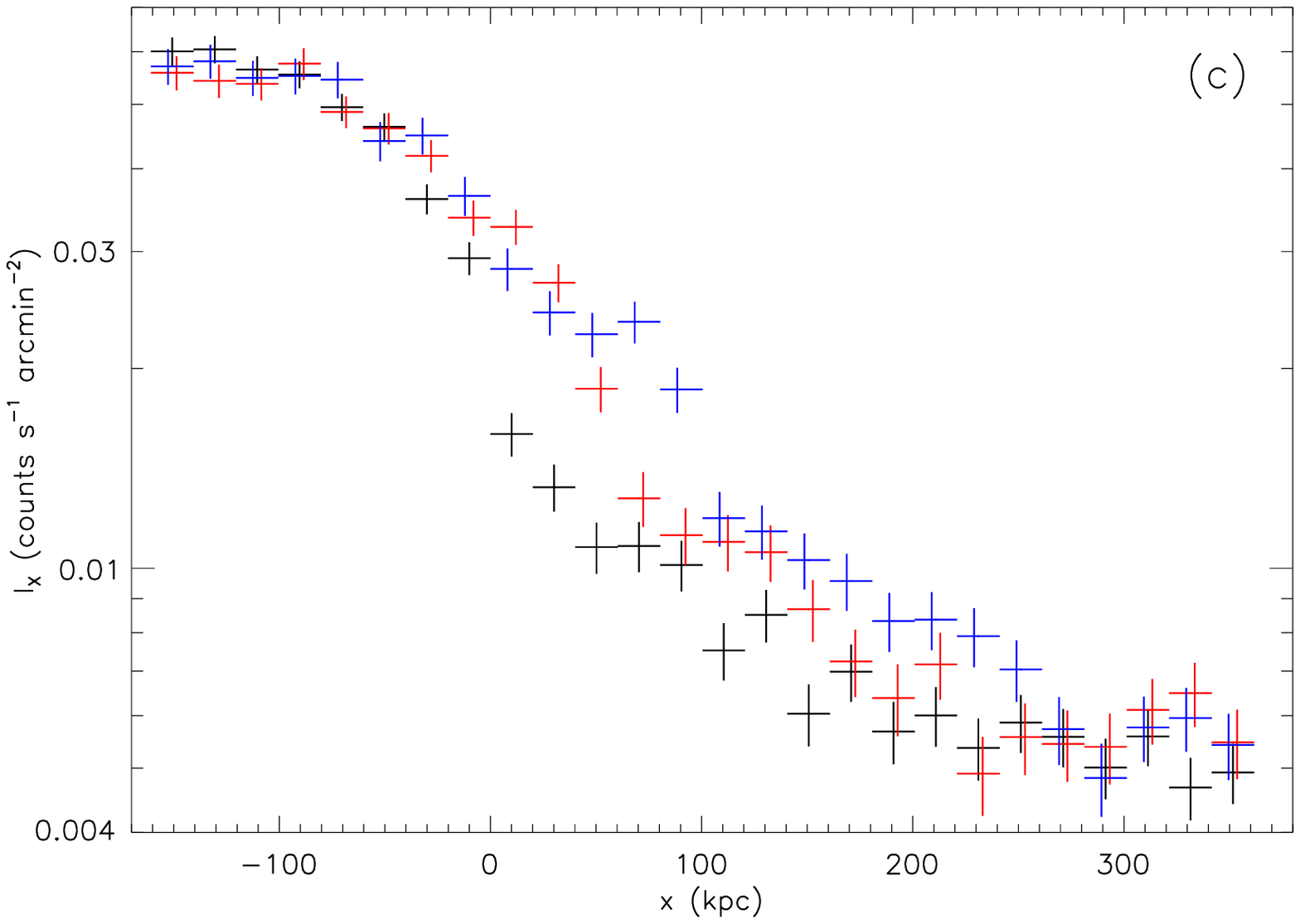}{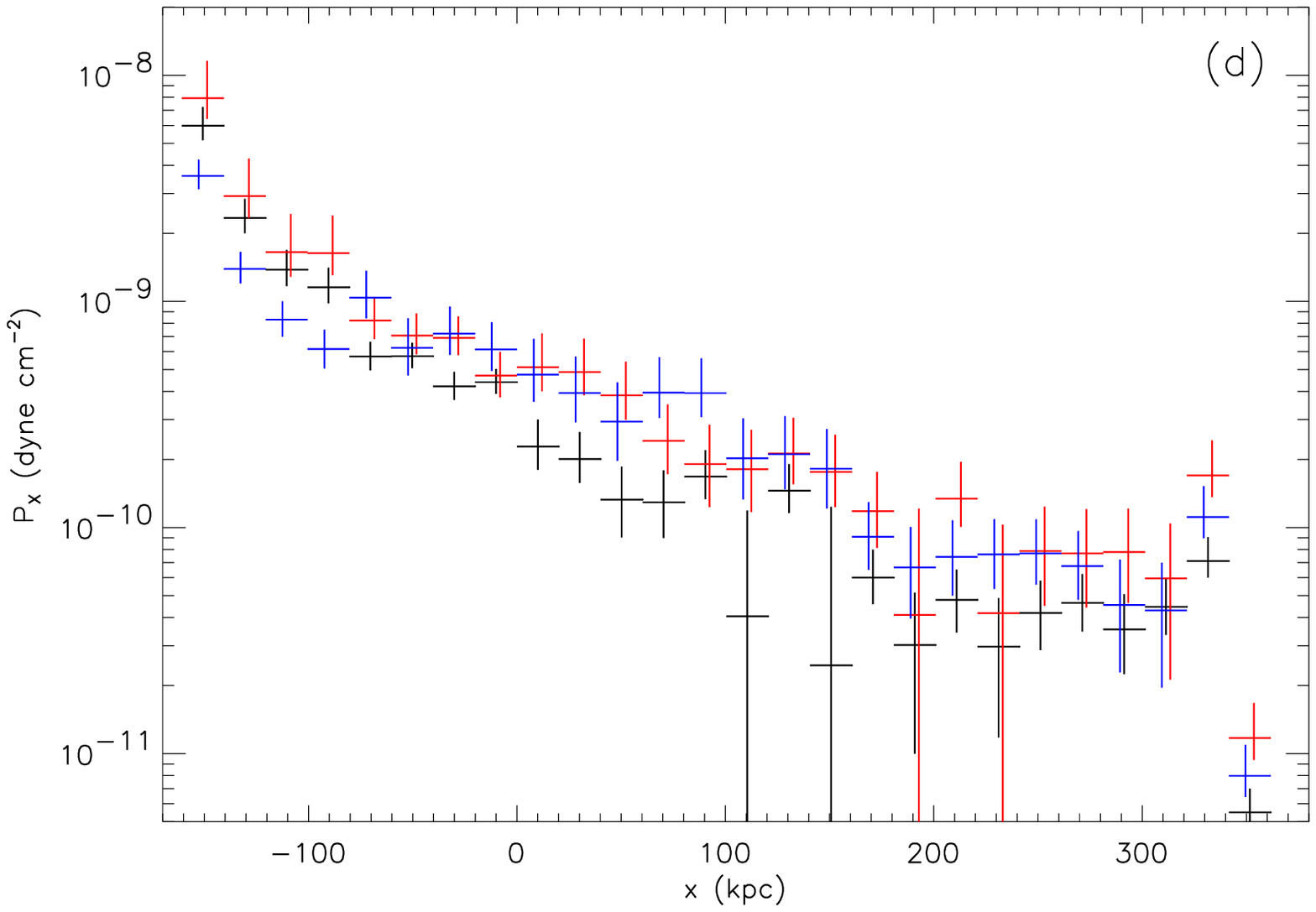}
\figcaption{{\it(a)} Residual emission from the cold front and associated
structure after subtraction of a $\beta$-model as in
Fig.~\ref{fig:2034_beta_resid}.
The wedges used for the analysis in \S\ref{sec:2034_cold_front} are
indicated by the dashed lines.
Projected {\it(b)}~temperature and {\it(c)}~surface brightness profiles,
and {\it(d)}~deprojected pressure profile across the cold front.
The three sectors, from east to west, are indicated in black, red, and
blue, respectively.
The red and blue error bars have been shifted slightly for clarity.
The point $x=0$ corresponds to the cold front (the dashed arc in {\it a}),
while negative radii are toward the cluster center and positive radii
extend out of the cluster.
\label{fig:2034_cold_front}}
\end{figure*}

Alternatively,
we consider the possibility that the south excess is a cluster
at a considerably larger redshift than that of Abell~2034.
By requiring that this cluster fit the $L_X$--$T$ relation of \citet{mar98},
it must be at a redshift of $0.30 < z < 1.25$.
Unfortunately, there are no strong line features in the X-ray spectrum of
the south excess region, so it is not possible to determine the redshift
associated with this emission directly from the spectrum.
There are only a handful of faint galaxies visible in the Digital
Sky Survey 2 (DSS-2) red image in the region of the south excess.
This would be consistent with the south excess being associated with
a moderately distant cluster.
Assuming the brightest cluster galaxy had an absolute magnitude
of $M_V = -22$, it would have an apparent R magnitude $m_R =
19.1$ at a  redshift of $z = 0.3$, and would be only slightly above the
sensitivity limit of the DSS-2.
The bulk of the galaxies in the cluster would therefore be below the
limit of the DSS-2.
\citetext{We assume a rest frame color $(V-R) = 0.9$ and take
$K$-corrections from \citealt*{cww80}.}
We have obtained moderately deep optical and near infrared images of the
field, which will be described in a subsequent paper, along with a more
detailed analysis of the X-ray emission from this region.
Even deeper optical imaging and follow-up spectroscopy should also be done
to verify the existence and proposed redshift of this background cluster.

\section{Northern Cold Front}
\label{sec:2034_cold_front}

To the north, the cluster contains a sharp surface brightness
discontinuity which was visible in an earlier {\it ROSAT}\/ image
(Fig.~\ref{fig:2034_rosat}).
As with other such features seen prior to {\it Chandra}, it was initially
assumed to be a merger shock.
Also like some other features of its kind, it in fact has turned out to be a
``cold front'' of the type first discovered in Abell~2142 \citep{mpn+00},
although the temperature contrast is small relative to Abell~2142 or other
similar cases, and the front itself does not appear to be stable.
In addition, the density contrast across the front is lower than in other
clusters: it is at most a factor of 2, compared to factors of 2--4 observed
in other cold fronts \citetext{e.g. A2142, \citealp{mpn+00}; A3667,
\citealp{vmm01b}}.
The cold front in Abell~2034 may also be less regular than those observed
in other clusters.
The temperature contrast across the front is also low: the gas inside the
surface brightness edge is not significantly cooler than that outside the
edge.
We conclude therefore that it is a cold front, since the high surface
brightness side of the edge is hot hotter than the low surface brightness
side, and therefore it cannot be a merger shock.
As can be seen in Figure~\ref{fig:2034_cold_front}a, there is a region of
excess emission ahead of the cold front to the north and northwest of the
front.
Moreover, the gas inside the center of the cold front is hotter than the
rest of the gas inside the front (see Figure~\ref{fig:2034_cold_front}b),
a further indication that the front may be unstable.
The leading edge of the cold front is roughly circular within
$\sim$$50^\circ$ of its axis of symmetry.
The axis of symmetry is at a position angle of approximately $2^\circ$
clockwise from North.
The cD galaxy associated with the cold front is ahead of the front,
$12^\circ$ west of the axis of symmetry.
The projected direction of motion, then, is probably somewhere between
these two values.

We extracted surface brightness and temperature measurements from a set of
concentric circular annular wedges which were defined to match the
curvature of the cold front as well as possible (see
Figure~\ref{fig:2034_cold_front}a).
These annuli were divided into three sectors so that the boundaries of the
sectors bracketed the hot region in the center of the front.
The projected temperature and surface bright profiles are shown in
Figure~\ref{fig:2034_cold_front}b,c.
We then deprojected the surface brightness profile separately for each
sector under the assumption of spherical symmetry.
The low surface brightness of the gas in these regions serves to minimize
projection effects in the temperature profile, so the differences between
the projected temperature profile and a deprojected one would be minimal.
We therefore used the projected profile for simplicity.
The deprojected surface brightnesses were then converted to electron
densities, which, when combined with the measured projected temperatures,
produced pressure profiles in each of the three sectors
(Figure~\ref{fig:2034_cold_front}d).
As mentioned above, the density contrast across the front is less than
a factor of two.
The pressures are consistent with being continuous across the cold front in
each sector, and remain continuous out to large radii.
Both within the cold front and at large radii past the cD galaxy, the
pressures are the same in each sector to within the errors.
Just outside the cold front, however, the pressure in the two sectors
containing the excess emission is higher than the pressure in the sector
without the excess, although not by more than 2-$\sigma$.
In all three sectors, the pressure drops with increasing distance from the
cluster center.

The overpressure in the region of the excess emission, combined with the
surprisingly hot gas in the central sector lead us to believe that the
front is unstable.
While the overpressure is measured at a fairly low significance, the
difference in temperatures between the sectors inside the front is
significant at greater than 90\% confidence.
Therefore, we consider a mechanism for creating an instability in the front.
As we discussed in \S\ref{sec:2034_image}, the cD galaxy associated with
the cold front has moved ahead of the front, as it has not been slowed by
the ram pressure forces that affect the ICM.
Since the dark matter is also presumably collisionless, it too has moved
ahead of the cold front.
This creates a situation in which the gradient of the potential is inverted
relative to the density and pressure gradients in the gas that used to be
at the center of the potential.
Thus, the dense gas which was formerly at the center of the cluster is now
further out from the center of the potential than is the less dense gas.
In the absence of a gravitational potential inside the front, a
Rayleigh-Taylor instability should form along the front \citep*[e.g.][]{jrt96}.
Instabilities should be visible if the timescale for their formation is short
compared to the timescale of the merger.
The timescale for the formation of this instability is dependent on the
acceleration due to the potential.
Unfortunately, since the dark matter and the X-ray emitting gas are
decoupled, it is not possible to reliably measure the potential, for
example using the technique described in \citet{vm02}

A further test of the stability of the cold front comes from a measurement
of its specific entropy.
In a stable cold front centered on the gravitational potential, the gas in
the center of the cold core has a lower specific entropy than the
surrounding gas.
Therefore, a higher specific entropy in the center of the cold front
would indicate that cold core has been disrupted.
The specific entropy is defined as
\begin{equation}
\label{eq:2034_delta_s}
\Delta s \equiv s_1 - s_2 = \frac{3}{2} k
\ln \left[
\left( \frac{T_1}{T_2} \right)
\left( \frac{\rho_1}{\rho_2} \right)^{-2/3}
\right] \, ,
\end{equation}
where the subscripts denote the two regions of interest.
Since this estimate of the entropy is only useful in a qualitative sense,
we approximate $\rho_1/\rho_2$ as $(I_{X1}/I_{X2})^{1/2}$, where $I_X$ is
the X-ray surface brightness in a given region.
We find that the hot region has a higher specific entropy than the
surrounding gas, with $\Delta s / (\frac{3}{2}k) = 0.6\pm0.2$.
Disruption of the core by Rayleigh-Taylor instability could induce
turbulence in the gas, which could explain the heating and consequent
increase in entropy.

We should note that hydrodynamic instability is by no means the only
possible explanation for the increase in entropy at the center of the
front.
A hydrodynamic shock, such as a merger shock, increases the specific
entropy of the shocked gas.
It is possible that a small merger shock has heated the gas at the center
of the front and left the rest of it untouched, thereby increasing the
entropy only in this central region.

Another possible explanation for the heating is that a now-extinct AGN
associated with the cD or one of the other nearby galaxies had heated this
gas at some point in the past.
We find this explanation to be quite unlikely, however, as there is no
evidence for radio activity in any of these galaxies, nor is there any
evidence for diffuse emission on the scale of the hot region, even at
low frequencies.
The radio relic, discussed in the next section, covers a much larger area,
including the cold gas to either side of the hot region.

\section{Radio Relic}
\label{sec:2034_relic}

\citet{ks01} reported a tentative detection of a radio relic in Abell 2034
near the position of the cold front.
The radio emitting electrons that produce relics are also expected to
produce X-rays by Inverse Compton (IC) scattering off of Cosmic Microwave
Background (CMB) photons.
The ratio of radio flux to IC flux is equal to the ratio of the energy
density in the magnetic field to that in the CMB.
Measuring both the radio synchrotron and X-ray IC emission would allow
both the magnetic field strength and the energy density in relativistic
electrons to be determined.

To measure the strength of Inverse Compton emission, we have extracted the
X-ray spectrum from the region of the radio relic, and fit it with the sum
of a MEKAL model and a power-law model, with the photon index of the
power-law set to $1+|\alpha|$, where $\alpha$ is the relic's spectral index
as determined by \citet{ks01}.
This analysis is complicated by the temperature structure discussed in
\S\ref{sec:2034_cold_front}.
To account for this, we extracted spectra separately from the regions having
different temperatures and fit their temperatures separately while requiring
a single power law model for the different regions.
The spectra are extremely well fit by a single-temperature MEKAL model,
with $\chi^2/{\rm d.o.f.} = 0.99$.
The addition of a power-law component improves the fit marginally.
The power-law component has a 1--10 keV flux of $7.5^{+4.9}_{-6.4}
\times10^{-14}$ ergs cm$^{-2}$ s$^{-1}$, which implies a mean
amplitude of the magnetic field of $0.3 < B < 0.9$ $\mu$G (90\% confidence),
including the 1-$\sigma$ measurement errors in the radio flux
\citep{ks01}.
This is consistent with the magnetic field strengths derived from analyses
of radio halos in other clusters \citep[e.g.][]{fg96, gfg+01}.
The errors on this measurement are admittedly quite large, but improved
calibration of the {\it Chandra}\/ detectors at energies below 1 keV could
reduce the errors on this measurement in a future re-analysis of the data.

\section{Summary}
\label{sec:2034_summary}

Our analysis of the {\it Chandra}\/ observation of the massive, moderate
redshift cluster Abell~2034 has revealed a number of interesting
features.
We have determined that the temperature of the cluster is fairly
constant out to a radius of at least about 800~kpc.
This includes the central region of the cluster, which shows no evidence
for the large cooling flow that had previously been claimed.
In contrast, the cluster shows some strong signatures of an ongoing
merger.
We have shown that the surface brightness discontinuity on the NE edge
of the cluster consistent with a cold front which is in the process of
being disrupted by a combination of ram pressure and the decoupling of
the gas from the dark matter potential.
This decoupling of the gas from the dark matter has also been demonstrated
to have occurred in 1E0657-56 \citep{mgd+02}.
A large concentration of galaxies, including a cD galaxy, is visible
just ahead of the cold front.
We suggest that these galaxies, including a cD galaxy, are centered
on the potential well of a subcluster, and that the gas in the cold front
was the cooling core of this subcluster.
The collisionless galaxies and dark matter have now moved ahead of
the collisional gas, which has been slowed by ram pressure during the
merger.
This has caused the front to begin to develop Rayleigh-Taylor
instabilities.
In addition to the cold front and the lack of a centrally condensed cooling
flow, other major evidence for the cluster being out of equilibrium
includes the significant offset of the central cD from the center of the
X-ray emission.

We have made a weakly constrained detection of IC emission from the
region of the proposed radio relic.
The magnetic field derived from this limit is consistent with other
measurements of cluster magnetic fields using radio halos.

There is a region of excess emission at a cooler temperature (4.9 keV
at the redshift of Abell~2034) to the south of the main cluster.
There is no clear evidence for any interaction between this gas and
the main Abell~2034 cluster.
The south excess is too faint to be an undisturbed cluster at the redshift
of Abell~2034 or closer, if it follows the normal $L_X$--$T$ relation.
It could be the remnant of a shocked subcluster, but this too would be too
faint, as shock heating would boost its luminosity by an amount that would
keep it more or less consistent with the normal $L_X$--$T$ relation.
Instead, we suggest that the south excess is a background cluster at an
estimated redshift of $0.3 < z < 1.25$.
Much more work is needed to confirm both the existence of and distance to
this new cluster, including deep optical imaging and spectroscopy.
We have begun this follow-up work, which we will discuss in a future paper.

\acknowledgements
Support for this work was provided by the National Aeronautics and Space
Administration through {\it Chandra}\/ Award Numbers
GO1-2123X,
GO2-3159X,
and
GO2-3164X
issued by the {\it Chandra}\/ X-ray Observatory Center,
which is operated by the Smithsonian Astrophysical Observatory for and on
behalf of NASA under contract NAS8-39073;
and by NASA contract NAS8-39073.
We thank John Houck for the S-Lang code which served as the basis for the
program to make the temperature map.
We also thank Roger Chevalier, William Forman, and D. Mark Whittle for
helpful comments on the manuscript.

\end{document}